\documentclass[conference]{IEEEtran}
\IEEEoverridecommandlockouts
\usepackage{cite}
\usepackage{amsmath,amssymb,amsfonts}
\usepackage{graphicx}
\usepackage{textcomp}
\usepackage{xcolor}
\usepackage{algorithm}
\usepackage{algpseudocode}
\usepackage[pagebackref,breaklinks,colorlinks]{hyperref}
\usepackage{xr}

\def\BibTeX{{\rm B\kern-.05em{\sc i\kern-.025em b}\kern-.08em
    T\kern-.1667em\lower.7ex\hbox{E}\kern-.125emX}}

\usepackage{balance}
\usepackage{url}
\usepackage{tikz}
\usepackage{pgfplots}
\usepackage{listings}
\usepackage{xcolor}
\usepackage[most]{tcolorbox}
\usepackage{tabularx}
\usepackage{multirow}
\usepackage[nameinlink]{cleveref} %
\usepackage[labelformat=simple,labelfont=bf]{subcaption} %
\usepackage{colortbl}
\usepackage{enumitem}
\usepackage{amsthm}
\usepackage[binary-units]{siunitx}
\usepackage{booktabs}
\usepackage{hyperref}
\theoremstyle{definition}

\usetikzlibrary{arrows, arrows.meta, tikzmark, ext.paths.ortho, matrix, positioning, shapes.geometric, calc, decorations.pathreplacing, calligraphy}
\usepgfplotslibrary{groupplots}

\crefname{section}{Section}{Sections}
\crefname{equation}{Equation}{Equations}
\crefname{figure}{Figure}{Figures}
\crefname{table}{Table}{Tables}
\crefname{listing}{Listing}{Listings}
\crefname{appendix}{Appendix}{Appendices}

\renewcommand{\figurename}{\textbf{Figure}}
\renewcommand{\tablename}{\textbf{Table}}
\renewcommand{\cite}[1]{~[\citenum{#1}]}

\definecolor{gray90}{gray}{0.9} %
\setlist[itemize]{leftmargin=10pt,itemindent=0pt,topsep=2pt,partopsep=2.5pt,parsep=1pt,itemsep=1.5pt,listparindent=\parindent{}}
\setlist[enumerate]{leftmargin=16pt,itemindent=0pt,topsep=2pt,partopsep=2.5pt,parsep=1pt,itemsep=1.5pt,listparindent=\parindent{}}

\newcommand{\nocolorref}[1]{\hypersetup{hidelinks}{\ref*{#1}}}

\pgfplotsset{compat=1.18}

\usepgfplotslibrary{groupplots}
\usetikzlibrary{decorations.pathmorphing}
\begin{document}
% --- PGFPlotsのバージョン互換性のための設定 ---
% これがないと、バージョンによってはエラーが出ることがあります
\pgfplotsset{compat=1.18}
\title{FlashGMM: Fast Gaussian Mixture Entropy Model for Learned Image Compression\\
\thanks{}
}

\author{\IEEEauthorblockN{Shimon Murai}
\IEEEauthorblockA{\textit{School of Fundamental} \\
\textit{Science and Engineering} \\
\textit{Waseda University}\\
Tokyo, Japan \\
octachoron@suou.waseda.jp}
% rewrite that fucking shit, it's lame and confusing
\and
\IEEEauthorblockN{Fangzheng Lin}
\IEEEauthorblockA{\textit{School of Engineering} \\
\textit{Institute of Science Tokyo}\\
Tokyo, Japan \\
lin.f.f849@m.isct.ac.jp}
\and
\IEEEauthorblockN{Jiro Katto}
\IEEEauthorblockA{
\textit{School of Fundamental} \\
\textit{Science and Engineering} \\
\textit{Waseda University}\\
Tokyo, Japan \\
katto@waseda.jp}}
\maketitle

\begin{abstract}
High-performance learned image compression codecs require flexible probability models to fit latent representations. Gaussian Mixture Models (GMMs) were proposed to satisfy this demand, but suffer from a significant runtime performance bottleneck due to the large Cumulative Distribution Function (CDF) tables that must be built for rANS coding. This paper introduces a fast coding algorithm that entirely eliminates this bottleneck. By leveraging the CDF's monotonic property, our decoder performs a dynamic binary search to find the correct symbol, eliminating the need for costly table construction and lookup. Aided by SIMD optimizations and numerical approximations, our approach accelerates the GMM entropy coding process by up to approximately 90x without compromising rate-distortion performance, significantly improving the practicality of GMM-based codecs. The implementation will be made publicly available at \url{https://github.com/tokkiwa/FlashGMM}.
\end{abstract}


\section{Introduction}

Learned image compression (LIC) has recently achieved state-of-the-art performance, surpassing traditional codecs such as VVC~\cite{jiang_mlic_2023, tcm}. LIC frameworks compress images into latent representations, whose probability distributions need to be modeled for entropy coding. The accuracy of this probability model is critical to the compression rate.

The Gaussian Mixture Model (GMM) \cite{cheng2020} has been demonstrated to be a highly effective probability model, significantly enhancing rate-distortion performance by accurately capturing the complex distributions of latent data. Despite its effectiveness, the GMM suffers from high computational complexity, which has hindered its adoption. This has led many recent works~\cite{minnen2020, jiang_mlic_2023, tcm} to favor simpler Gaussian Single Models (GSM), thereby forgoing the performance gains offered by GMMs.

This work addresses the critical challenge of GMM's computational cost. We propose a novel and highly efficient algorithm, named \textbf{FlashGMM}, for entropy coding with GMMs. Traditional implementations for GMM-based coding~\cite{cheng2020, lin_multistage_2023, begaintCompressAIPyTorchLibrary2020} rely on generating a quantized CDF table prior to encoding and decoding. This table generation process is very computationally heavy, thus a major performance bottleneck. Although the GMM-based entropy model has evolved from its original single-threaded implementation~\cite{cheng2020} to a parallelized GPU-accelerated version in CompressAI~\cite{begaintCompressAIPyTorchLibrary2020}, and was further optimized with the PyTorch C++ API by Lin et al.~\cite{lin_multistage_2023} for efficient data transfer, the fundamental probability estimation algorithm remains unchanged, as detailed in Section~\ref{sec:prev_gmm_algorithm}. 

Our method fundamentally redesigns the calculation process. By exploiting the monotonic property of the CDF, we employ a binary search algorithm to find the inverse CDF on the fly during decoding, completely obviating the need for a pre-computed table. To further boost the speed, we integrate a fast approximation for the Gaussian CDF and leverage Single Instruction, Multiple Data (SIMD) extensions to parallelize the computations across mixture components. Our experiments show that FlashGMM is approximately 90 times faster than the CompressAI GMM implementation and more than 10 times faster than the optimized variant by Lin et al.~\cite{lin_multistage_2023}, and even surpasses the performance of GSM, all without any tradeoff in rate-distortion.

\begin{figure}[t]
\scalebox{0.90}{
\begin{tikzpicture}%[trim axis left, trim axis right]

\begin{groupplot}[
name=overheadfig,
scaled ticks=false,
y axis line style={draw=none},
ymajorgrids=true, yminorgrids=true, minor tick num=1,
major grid style={line width=.6pt,draw=gray!60},
ybar=0pt, xtick=data,
width=0.48\textwidth, height=6cm,
yticklabel={\pgfmathprintnumber{\tick}},
% --- 変更点 1: symbolic x coords のラベルを {} で囲み、\\ で改行 ---
symbolic x coords={{GMM\\(CompressAI)}, {GMM\\(Lin2023)}, FlashGMM, GSM},
% --- 変更点 2: xticklabel style に align を追加し、改行を有効化 ---
xticklabel style={align=center, yshift=1.5mm},
xtick style={draw=none}, x axis line style={line width=1pt,draw=black},
legend style={at={(1, 1)}, anchor=north east, font=\footnotesize},
legend image code/.code={\draw [#1] (0cm,-0.1cm) rectangle (0.2cm,0.15cm); },
legend cell align={left},
legend to name=overheadlegend_exp_runtime,
legend columns=-1,
group style={
group size=1 by 2,
xticklabels at=edge bottom,
vertical sep=-0.4mm,
},
set layers,
]
\nextgroupplot[
ymin=1.5,ymax=2.3,
ytick distance=1,
x axis line style={draw=none},
height=3.5cm,
bar width=.3cm,
enlarge x limits=0.15,
y filter/.code={\pgfmathifthenelse{#1 < 0.01}{NaN}{#1},},
]

\begin{scope}
\clip (axis cs: {[normalized]-1}, 1.0) rectangle (axis cs: {[normalized] 5}, 5);
\addplot[fill=black!75]
%all enc
coordinates {
({GMM\\(CompressAI)}, 2.020191209)
({GMM\\(Lin2023)}, 1.)
(FlashGMM, 1.)
(GSM, 1.)
};

\addplot[fill=black!50]
%all dec
coordinates {
({GMM\\(CompressAI)}, 2.018483147) % 座標指定も新しいラベル名に合わせる
({GMM\\(Lin2023)}, 1.)
(FlashGMM, 1.)
(GSM, 1.)
};
\addplot[fill=black!25]
%GMM encode
coordinates {
({GMM\\(CompressAI)}, 1.964887038) % 座標指定も新しいラベル名に合わせる
({GMM\\(Lin2023)}, 1.)
(FlashGMM, 1.)
(GSM, 1.)
};
\addplot[fill=black!5]
%GMM decode
coordinates {
({GMM\\(CompressAI)}, 1.991) % 座標指定も新しいラベル名に合わせる
({GMM\\(Lin2023)}, 1.)
(FlashGMM, 1.)
(GSM, 1.)
};

\end{scope}
\begin{pgfonlayer}{axis foreground}
\draw [decorate, decoration={snake, amplitude=0.8mm, segment length=7mm}, line width=1.5mm, black] (axis cs: {[normalized]-0.5}, 1.5) -- (axis cs: {[normalized] 3.5}, 1.5);
\draw [decorate, decoration={snake, amplitude=0.8mm, segment length=7mm}, line width=1mm, white] (axis cs: {[normalized]-0.5}, 1.5) -- (axis cs: {[normalized] 3.5}, 1.5);
\end{pgfonlayer}

\nextgroupplot[
ymin=0,ymax=0.34,
ytick distance=0.1,
axis x line*=bottom,
height=3.5cm,
bar width=.3cm,
enlarge x limits=0.15,
ylabel={Execution Time (sec)},
ylabel style={xshift=0.5cm, yshift=0cm}
]
\addplot[fill=black!75]
%ALL enc
coordinates {
({GMM\\(CompressAI)},2.020191209
) % 座標指定も新しいラベル名に合わせる
({GMM\\(Lin2023)},0.270657348
)
(FlashGMM, 0.05624096
)
(GSM, 0.065469846
 )
};

\addplot[fill=black!50]
%ALL dec
coordinates {
({GMM\\(CompressAI)},2.018483147) 
({GMM\\(Lin2023)},0.26240793)
(FlashGMM,0.03964019)
(GSM,0.047761905)
};
\addplot[fill=black!25]
%GMM enc
coordinates {
({GMM\\(CompressAI)}, 1.965) % 座標指定も新しいラベル名に合わせる
({GMM\\(Lin2023)}, 0.2397 )
(FlashGMM, 0.02249)
(GSM, 0.0290 )
};

\addplot[fill=black!5]
%GMM dec
coordinates {
({GMM\\(CompressAI)}, 1.991) % 座標指定も新しいラベル名に合わせる
({GMM\\(Lin2023)}, 0.2565)
(FlashGMM, 0.03314)
(GSM, 0.0386 )
};

\legend{Total compress, Total decompress, Entropy encode, Entropy decode}
\end{groupplot}

\node[below, xshift=3cm, yshift=2.5cm, text=black] {\nocolorref{overheadlegend_exp_runtime}};

\end{tikzpicture}
}
%\vspace{-3mm}
\caption{Execution time of the different algorithms including GMM (CompressAI)\cite{begaintCompressAIPyTorchLibrary2020}, GMM (Lin2023)\cite{lin_multistage_2023}, FlashGMM (ours), and GSM\cite{ballé2018variational}. The left two bars indicate the model's total compression and decompression times, and the right two bars represent only the entropy coding time, excluding the main transform and the hyperprior network. Complete results are in Table~\ref{tab:runtime_comparison}.}
\label{fig:exp-runtime-overhead}
%\vspace{-3mm}
\end{figure}
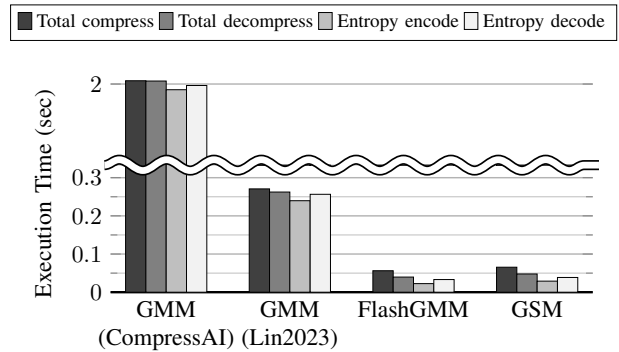

% omitting. too long

% Our contributions are summarized as follows:
% \begin{enumerate}
% \item We propose a novel, fast and paralellized algorithm for computing the GMM-based entropy model in learned image compression.
% \item We provide a comprehensive performance comparison, demonstrating that our proposed method is not only faster than, but also achieves a better rate-distortion performance than the commonly used Gaussian Single Model(GSM).
% \item We will release our implementation publicly to facilitate its adoption into future image compression models.
% \end{enumerate}
\pgfplotsset{compat=1.18} % 互換性バージョンを指定します
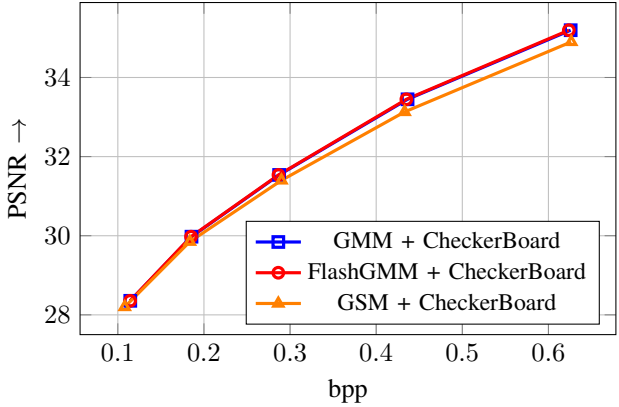
\begin{figure}[t]
\begin{tikzpicture}
\begin{axis}[
    xlabel={bpp},
    table/col sep=comma,
    ylabel={PSNR $\rightarrow$},
    legend pos=south east,
    grid=major,
    height=6cm,
    width=0.48\textwidth,
    scaled ticks = false,
    tick label style={/pgf/number format/fixed},
    legend style={font=\small},
]

% CSVファイル1のデータをプロット
\addplot[
    color=blue,
    mark=square,
    very thick,
] table [x index=0, y index=1, col sep=comma, header=false] {csv/naivegmm.csv};
\addlegendentry{GMM + CheckerBoard}

% CSVファイル3のデータをプロット
\addplot[
    color=red,
    mark=o,
        very thick,
] table [x index=0, y index=1, col sep=comma, header=false] {csv/faster_logi.csv};
\addlegendentry{FlashGMM + CheckerBoard}

% CSVファイル2のデータをプロット
\addplot[
    color=orange,
    mark=triangle,
        very thick,
] table [x index=0, y index=1, col sep=comma, header=false] {csv/gsm.csv};
\addlegendentry{GSM + CheckerBoard}

% % CSVファイル4のデータをプロット
% \addplot[
%     color=purple,
%     mark=star,
%         very thick,
% ] table [x index=0, y index=1, col sep=comma, header=false] {csv_lpips/hific_lpips.csv};
% \addlegendentry{HiFiC[12]}

% \addplot[
%     color=red,
%     mark=diamond,
%         very thick,
% ] table [x index=0, y index=1, col sep=comma, header=false] {csv_lpips/ours_lpips.csv};
\addlegendentry{ours}
\end{axis}
\end{tikzpicture}
\caption{Rate-Distortion Curve of different entropy coding methods. Models are based on Checkerboard\cite{heCheckerboardContextModel2021b} and evaluated on Kodak\cite{kodak} dataset. For GMM and GSM, we adopted the CompressAI\cite{begaintCompressAIPyTorchLibrary2020} implementation.}
\label{fig:rdcurve}
\end{figure}
\section{Related Work}

\subsection{Learned Image Compression with Gaussian Entropy Model}

The basic structure of learned image compression is proposed by~\cite{balle2017}. First, the input image $\boldsymbol{x}$ is transformed to a latent representation $\boldsymbol{y}$ with some learnable transform $g_a(\cdot)$. The latent representation $\boldsymbol{y}$ is then quantized to $\hat{\boldsymbol{y}}$. We assume some factorized distribution $p(\boldsymbol{y}|\boldsymbol{\theta}) = \prod_i p(y_i | \boldsymbol{\theta}) $ for $\boldsymbol{y}$, and predict the conditional parameters $\boldsymbol{\theta}$ with a subnetwork called hyperprior~\cite{ballé2018variational}. 
%Also, due to noised quantization \cite{balle2017}, the actual modeling is the "noised" distribution expressed as
Noise is added during training to simulate quantization~\cite{balle2017} such that
\begin{align} p(\hat{y}_i) = c(\hat{y}_i + 0.5) - c(\hat{y}_i - 0.5) \label{eq:noised_quantized_distribution_modeling}\end{align} where $c(\cdot)$ is the cumulative distribution function (CDF) of the probability model of $y$. 

In Gaussian Single Models, we assume $p(y_i|\boldsymbol{\theta}) = \mathcal{N}(y_i | \mu_i, \sigma_i) $. Hereafter, let $\Phi(\cdot|\mu, \sigma)$ be the CDF of normal distribution conditioned by the mean $\mu$ and standard deviation $\sigma$. When we write $\Phi(\cdot)$ without the means and standard deviations, that indicates the CDF of standard normal distribution $\mathcal{N}(0,1)$. Then the distribution of the quantized latent can be written as $p(\hat{y}|\boldsymbol{\theta}) = \Phi(\hat{y_i} +0.5|\mu, \sigma) - \Phi(\hat{y_i} - 0.5|\mu, \sigma)$. 

The Gaussian Mixture Model (GMM) proposed by~\cite{cheng2020} further extends this modeling to the mixture of $K$ Gaussians. The GMM assumes the following distribution:
\begin{align}
    p(y_i|\boldsymbol{\theta}) = \sum_{k=1}^Kw_{i,k}\mathcal{N}(y_{i} | \mu_{i,k}, \sigma_{i.k}) 
\end{align}
where $w_{i,k}$ is a weight parameter that satisfies $\sum_{k=1}^K w_{i,k} = 1$. The use of GMM effectively models the complex distributions of latent variables, while trading off runtime performance due to the computation overhead.

\subsection{Asymmetric Numeral Systems}
Asymmetric Numeral Systems (ANS)~\cite{duda2009asymmetricnumeralsystems, duda2014asymmetricnumeralsystemsentropy} is a widely used entropy coding scheme in modern learned image compression, valued for its excellent balance of speed and compression rate, and therefore adopted in many LIC implementations. Unlike arithmetic coding, ANS uses a single natural number as its state.

Its variant, \textbf{rANS}, approximates a symbol's probability $p_s$ as a rational number, $p_s \approx l_s / m$, where $m = \sum_s l_s$. Based on this, the encoding function that maps the current state $x$ and symbol $s$ to a new state $x'$ is defined as:
\begin{align} C(s, x) = m \cdot \lfloor x/l_s \rfloor + b_s + \mathrm{mod}(x, l_s), \end{align}
where $b_s$ is the cumulative frequency of preceding symbols. 

In the context of learned image compression, the rANS framework is implemented using the Cumulative Distribution Function (CDF). Both encoding and decoding operate solely on CDF values, without directly using the Probability Mass Function (PMF). This requires the decoder to perform an inverse mapping from a retrieved CDF value back to the original symbol. The inverse CDF of a normal distribution is analytically intractable: approximating this inverse numerically would be computationally expensive and susceptible to floating-point precision errors. This inverse calculation is therefore typically solved by creating and searching a quantized CDF lookup table, a technique used in seminal works such as~\cite{balle2017}.

Techniques have been employed to reduce the size of such lookup tables and even enable them to be preprocessed in implementations of Gaussian Single Models, such as the one in CompressAI~\cite{begaintCompressAIPyTorchLibrary2020}. The symbol to be encoded $s \sim \mathcal{N}(\mu, \sigma)$ is shifted by $\mu$, passing $s - \mu \sim \mathcal{N}(0, \sigma)$ to the encoder, therefore eliminating $\mu$ as a parameter in the distribution. As many of these normal distributions $\mathcal{N}(0, \sigma)$ (with trivial differences in $\sigma$) would look alike after quantization, CompressAI also quantizes the $\sigma$ into 64 levels, logarithmically evenly spaced between $0.11$ and $256$. Therefore, only 64 CDF lookup tables have to be built, and they can be pre-built as the 64 possible quantized $\sigma$ values are constant. This is the key to why Gaussian Single Models do not suffer from severe performance problems during entropy coding in major implementations.
\subsection{Existing GMM Algorithm}
\label{sec:prev_gmm_algorithm}

The fundamental performance challenge in building CDF lookup tables for GMM lies in the fact, that it is unable to be optimized in a way similar to the one applied to GSM above. Since GMM uses a combination of multiple normal distributions, more parameters determine the shape of the distribution (multiple $\mu_i$s and $\sigma_i$s vs. only one $\sigma$ in optimized GSM). The CDF tables are therefore impossible to be prebuilt, and can only be computed as needed. 

Here we review how the existing GMM algorithm based on the original~\cite{cheng2020} and the CompressAI's ~\cite{begaintCompressAIPyTorchLibrary2020} implementation. Let the \textit{input} $\hat{\boldsymbol{y}}$ be the quantized integer vector of size $|\hat{\boldsymbol{y}}|$ and $\boldsymbol{\mu}, \boldsymbol{\sigma}, \boldsymbol{w}$ are corresponding Gaussian parameters of size $(|\hat{\boldsymbol{y}}|, K)$. Let the \textit{alphabet} $\mathcal{Y} \subset \mathbb{Z}$ be the set of possible values where every element  $ y_i \in \hat{\boldsymbol{y}}$ satisfies $ y_i \in \mathcal{Y}$, and let $|\mathcal{Y}|$ denote the size of alphabet $\mathcal{Y}$. 

The previous model prepares $(|\hat{\boldsymbol{y}}|, |\mathcal{Y}|)$-sized table $\mathcal{T}$. The table contains the quantized CDF for all possible values $y_j \in \mathcal{Y}$. The $(i,j)$-th element of $\mathcal{T}$ can be represented as 

\begin{align}
\mathcal{T}_{i,j} &=\sum_{l=0}^j p(y_l|\mu_{i}, \sigma_{i} ,w_{i})
\\\ p(y_l|\mu_{i}, \sigma_{i} ,w_{i}) &= \sum_{k=1}^K w_{i,k}(\Phi(y_l + 0.5| \mu_{i,k}, \sigma_{i,k}) \\ &\space - \Phi(y_l - 0.5| \mu_{i,k}, \sigma_{i,k}))  \textrm{ where } \space y_l \in \mathcal{Y}.
\end{align}

It takes $O(K|\hat{\boldsymbol{y}}||\mathcal{Y}|)$ to construct the table. The table enables us to calculate CDF from PMF and guarantees encoder-decoder compatibility in exchange for heavy computational costs. After the symbols are encoded, the decoder needs to calculate the inverse cumulative distribution function $\text{CDF}^{-1}(d)$ where $d$ is the previously decoded CDF value. Given the shared table, the decoder can search over the column of the table $\mathcal{T}_{i}$ to find the correct value of $\hat{y}$. The overall theoretical complexity is  $O(K|\hat{\boldsymbol{y}}||\mathcal{Y}|)$ for both en/decoding. 

%\color{red}
%In conventional implementations, the pre-computation of a Cumulative Distribution Function (CDF) table was an indispensable step.

%First, the analytical computation of the CDF is intractable. The probability mass for a discrete symbol $\hat{y}$ defined in Eq.~\ref{eq:noised_quantized_distribution_modeling} contains $c(x)$, each involving an integral of a Gaussian probability density function (PDF). Therefore, it is challenging to integrate the PDF again to find a numerically accurate CDF function. Consequently, the only practical method was to construct a PMF table and then derive the CDF table by performing a cumulative sum across it.

%Second, the entropy decoding process requires inverse mapping of CDF. Similar to the forward computation, the inverse CDF of a GMM is analytically intractable. Approximating this inverse numerically would be computationally expensive and susceptible to floating-point precision errors, rendering it unsuitable for a high-speed decoder. The use of a pre-computed CDF table avoids this issue, allowing for an efficient inverse mapping through a table search algorithm.

%For these reasons, the table-based approach was a mandatory prerequisite for entropy coding when employing the conventional GMM formulation.\color{black}
This table construction is computationally intensive but highly parallel, and therefore CompressAI~\cite{begaintCompressAIPyTorchLibrary2020} and Lin et al.~\cite{lin_multistage_2023} accelerated this process by utilizing a GPU. However, we show that FlashGMM completely eliminates this computational intensity, outperforms the above implementations, and does not require the use of a GPU.
\begin{table*}[t]
\centering
\caption{Comparison of speed and performance for two different settings. Each of the experiments are based on the CheckerBoard\cite{heCheckerboardContextModel2021b} model. Note that $|\mathcal{S}|$ denotes the alphabet of scales and it additionally takes $O(|\mathcal{S}|)$ to initialize GSM. The speed columns contain encoding/decoding speed. The detailed experiment settings are shown in~\ref{sec:setting}.}
\label{tab:runtime_comparison}
\renewcommand{\arraystretch}{1.3} % Adjust row height for better readability
\resizebox{\textwidth}{!}{
\begin{tabular}{|l|c|c|c|c|c|c|}
\hline
\multirow{2}{*}{\textbf{Method}} & \multicolumn{5}{c|}{\textbf{Speed}} & \textbf{Performance} \\ \cline{2-7} 
 & \textbf{Theoretical} & \textbf{Total time, Normal} & \textbf{Coding time, Normal} & \textbf{Total Time, High} & \textbf{Coding time, High} & \textbf{BD-Rate (PSNR)} \\
\hline \hline
% NOTE: The O-notation for 'Theoretical' and the time values in each cell are placeholders.
%       Please replace them with your actual values.
GSM       & $O(|\hat{\boldsymbol{y}}| + |\mathcal{S}|) / O(|\hat{\boldsymbol{y}}| + |\mathcal{S}|)$ &65ms
 / 48ms
 &29ms
 / 39ms
 & 48ms / 39ms & 23ms / 31ms & 0.00\% (Anchor) \\ \hline
GMM(CompressAI)  & $O(K|\mathcal{Y}||\hat{\boldsymbol{y}}|) / O(K|\mathcal{Y}||\hat{\boldsymbol{y}}|)$ & 2020ms
 / 2018ms
 & 1965ms
 / 1991ms
 & 1907ms / 1910ms &  1864ms / 1890ms & -4.02\% \\ \hline
GMM(Lin2023) & $O(K|\mathcal{Y}||\hat{\boldsymbol{y}}|) / O(K|\mathcal{Y}||\hat{\boldsymbol{y}}|)$ & 271ms
 / 262ms &240ms
 / 256ms
& 189ms / 181ms & 148ms / 169ms
 & -4.02\% \\ \hline
FlashGMM & $O(K|\hat{\boldsymbol{y}}|) / O(|\hat{\boldsymbol{y}}| K\log |\mathcal{Y}|)$ & 56ms / 40ms & 22ms
 / 33ms
 & 42ms / 35ms & 19ms / 29ms & \textbf{-4.28\%} \\ \hline
\end{tabular}
}
\end{table*}
\section{Proposed Method}
\subsection{Faster Gaussian Mixture Algorithm}

Our goal is to eliminate CDF table construction during GMM calculation to achieve an efficient GMM algorithm. Here we leverage two properties:
\begin{itemize}
    \item rANS only requires quantized values of CDF at the symbol of interest for encoding/decoding.
    \item CDF is a monotonic function.
\end{itemize}

The second property enables us to calculate the inverse CDF function using a binary search. We simply model $\hat{y}_i \sim \sum_k \mathcal{N}(\hat{y}_i| \mu_{i,k}, \sigma_{i,k}) \cdot w_{i,k} $ while previous models assume $y \sim \sum_k \mathcal{N}(y_i| \mu_{i,k}, \sigma_{i,k}) \cdot w_{i,k} $ and compute the distribution after noised quantization $p(\hat{y}_i)$ using convolution as shown in Equation \ref{eq:noised_quantized_distribution_modeling}. 

Our encoding/decoding process is as follows: when the value $\hat{y}_i$ is sent to the encoder, it directly computes $\text{CDF}(y_i|\mu_i, \sigma_i, w_i)$ and encodes the symbol with rANS. However, the decoder needs to calculate the inverse CDF $\text{CDF}^{-1}(d)$ where $d$ is the previously decoded CDF value to retrieve the symbol. To satisfy this requirement, the decoder performs a binary search for the corresponding $\hat{y}$ within $\text{CDF}(\cdot|\mu_i, \sigma_i, w_i)$. Because of the monotonicity of the CDF, this operation is verified to fetch the correct value of $\hat{y}$.
%The calculation is done with 16 bit precision to avoid calculation error. 
The encoding takes only $O(K|\hat{\boldsymbol{y}}|)$ and the decoding takes $O(K|\hat{\boldsymbol{y}}|\log|\mathcal{Y}|)$.% where $K$ is neglisible because of SIMD parallelization.

This method guarantees that the encoder/decoder correctness is not affected by floating-point errors, as the series of calculations performed by the encoder/decoder are essentially on the same numerical function $\text{CDF}(\cdot|\mu_i, \sigma_i, w_i)$. Computing $\text{CDF}^{-1}$ numerically without using binary search is not only infeasible for GMM, but also may incur decoding errors when probabilities are close to symbol boundaries, whereas our method guarantees decode correctness.

\subsection{Fast Gaussian CDF Approximation}
\label{sec:approx_gmm}
Throughout the computation of the entropy model, the Gaussian Cumulative Distribution Function (CDF), $\Phi(\cdot)$, is a computationally expensive component. Its standard definition requires numerical integration of the probability density function:
\begin{align}
    \Phi(x) = \frac{1}{\sqrt{2\pi}} \int_{-\infty}^{x} \exp\left(\frac{-t^2}{2}\right) dt.
    \label{eq:cdf_def}
\end{align}
The complexity in calculating the CDF of normal distributions is problematic in many applications, and therefore several closed-form approximations were proposed~\cite{Plya1949RemarksOC, abramowitzHandbookMathematicalFunctions2013, hendrycksGaussianErrorLinear2023}. To circumvent this costly operation, we explore three candidates:

(1) the P\'olya approximation~\cite{Plya1949RemarksOC}, 
\begin{align}
    \Phi(x) \approx \frac{1}{2} \left( 1 + \sqrt{1 - \exp(-x^2)} \right)
\end{align}

(2) the Abramowitz \& Stegun (A\&S) approximation~\cite{abramowitzHandbookMathematicalFunctions2013},

\begin{align}\Phi(x) \approx 1 - Z(x)(b_1t + b_2t^2 + b_3t^3 + b_4t^4 + b_5t^5)\end{align}
where $Z(\cdot)$ is standard gaussian PDF, $t = 1/(1 + px)$ and the other variables are pre-computed fixed constant. 

(3) the logistic function approximation, which is used in the Gaussian Error Linear Unit (GELU)~\cite{hendrycksGaussianErrorLinear2023}. The logistic approximation is given by:
\begin{align}
    \Phi(x) \approx \frac{1}{1 + \exp(-1.702x)}.
    \label{eq:logistic_approx}
\end{align}
Considering the trade-off between accuracy and computational speed, we adopted the simplest logistic function approximation for our final model.

\subsection{SIMD Parallelization}
Although our approach skips the construction of a lookup table for the Cumulative Distribution Function (CDF), it still necessitates the computation of $K$ Gaussian components for each symbol.
Noting that these computations are mutually independent, we introduce \textit{Single Instruction, Multiple Data} (SIMD) parallelization to accelerate this process.
SIMD is a parallel processing paradigm that enables a single instruction to operate on multiple data points simultaneously, making it particularly effective for vectorized calculations, such as those in the Gaussian Mixture Model (GMM).
%Although the original implementation with error function (erf) is not supported by SIMD, 
The numerical approximations we use enable a single SIMD instruction to concurrently process maximum $K=4$ components under AVX2; most LIC GMM models use $K=3$, well within the capacity.
Consequently, the theoretical computational overhead compared to GSM is reduced to a negligible level.

\section{Experiment}
\subsection{Experiment Setting}
\label{sec:setting}
We adopted the Checkerboard model~\cite{heCheckerboardContextModel2021b} as our baseline, which was introduced as an enhancement to the original Cheng et al. model~\cite{cheng2020} and features a simplified two-pass context model. Utilizing the implementation in CompressAI~\cite{begaintCompressAIPyTorchLibrary2020}, we trained this model with two distinct entropy models: the Gaussian Single Model (GSM) and our proposed FlashGMM.

For training, we utilized the MLIC dataset~\cite{jiang_mlic_2023}, which comprises approximately 108,000 images. During training, each image was randomly cropped to $256 \times 256$ pixels, and the models were trained for 150 epochs with a batch size of 16. We employed the Adam optimizer~\cite{Kingma2014AdamAM}, initializing the learning rate to $1 \times 10^{-4}$ and subsequently reducing it by a factor of 10 at the 120, 140, and 145 epoch. The Lagrangian parameter $\lambda$ was set to $\lbrace0.0018, 0.0035, 0.0067, 0.013, 0.025\rbrace$. 

Our experiments were conducted in two distinct hardware environments to measure performance. The \textit{Normal} environment consisted of an Intel Core i7-7700 CPU and an NVIDIA RTX 2080 SUPER GPU. The \textit{High-speed} environment was equipped with an AMD Threadripper 3960X CPU and an NVIDIA RTX 3090 GPU. 

\subsection{Evaluation}
\subsubsection{RD Performance}
The Rate-Distortion (RD) curves obtained from our experiments are presented in Fig. \ref{fig:rdcurve}.
Furthermore, we calculate the Bj{\o}ntegaard Delta Rate (BD-Rate)~\cite{bjontegaard} using GSM as an anchor, with the results shown in Table \ref{tab:runtime_comparison}.
These findings indicate that our method not only exceeds the performance of the GSM but also slightly exceeds that of the conventional GMM. 
We attribute this improvement to the omission of noisy distribution modeling during inference, which we believe enhances the model's confidence and thus leads to improved performance. 
In fact, Table~\ref{tab:cdf_approx_comparison} demonstrates that performance is enhanced even when calculations are performed without approximation methods, further supporting our approach.
\subsubsection{Runtime Performance}

For our speed analysis, we utilized the Kodak image dataset~\cite{kodak}. 
The ``Encoding'' and ``Decoding'' times were measured by summing the processing time of the entropy model and the rANS coder during the compression and decompression stages of the Checkerboard model, respectively. 
Each process was executed 10 times, and the average time was recorded. 
The results for each method are summarized in Table~\ref{tab:runtime_comparison}.

Across all tests, our proposed method demonstrated the best performance. 
Our FlashGMM surpasses the performance of other GMM implementations, despite not requiring a GPU for the costly CDF-table construction process, which is eliminated. The freed GPU resources can contribute towards further FPS when implemented under more efficient architectures~\cite{lin_icip}. FlashGMM also outperforms GSM in terms of performance, suggesting that there may still be room for optimization in current GSM algorithms.

\begin{table}[t]
\centering
\caption{Performance comparison of Gaussian CDF approximation methods. The BD-Rate anchor is GSM.}
\label{tab:cdf_approx_comparison}
\begin{tabular}{l|c|c}
\hline
\textbf{Approximation Method} & \textbf{En/Decoding Speed} & \textbf{BD-Rate} \\
\hline \hline
No Approximation           & 56.5ms / 127ms & -4.24\% \\
Pólya            & 23.5ms / 34.3ms & -4.02\% \\
Abramowitz \& Stegun      & 22.5ms / 33.7ms & -4.24\% \\
Logistic                  & 22.5ms / 33.2ms & -4.28\% \\
\hline
\end{tabular}
\end{table}
\section{Ablation Study}
\label{sec:approx_comparison}

We conducted a comparative analysis of the speed and rate-distortion performance of the various Gaussian CDF approximation methods discussed in Section~\ref{sec:approx_gmm}. 
For the non-approximated baseline, calculations were performed using the \texttt{std::erfc} function from the C++ standard library without SIMD parallelization. 
All other approximation methods were implemented using SIMD instructions.

The results indicate that the baseline (``no approximation'') and the A\&S approximation achieve comparable performance. 
This is likely because the accuracy of A\&S approximation is sufficient for our application, as its maximum error is shown to be below $7.5\times 10^{-8}$ (See Sec. 26.2.17 of~\cite{abramowitzHandbookMathematicalFunctions2013}). 
Interestingly, the logistic approximation yielded even better rate-distortion performance. 
We hypothesize that this is because the true latent distribution is not perfectly Gaussian, and the logistic function coincidentally provides a more accurate fit.

Based on these findings, we recommend using the logistic approximation for optimal performance. 
Alternatively, the A\&S approximation can be used as a more conservative and robust choice for practical applications.

\section{Conclusion}

In this paper, we addressed the significant computational bottleneck of Gaussian Mixture Models in learned image compression. 
We introduced FlashGMM, a novel search-based coding algorithm that completely eliminates the need for pre-computed CDF tables. 
By leveraging on-the-fly CDF calculations with SIMD parallelization and fast numerical approximations, our method achieves a speedup of up to 90x compared to existing methods. 
Crucially, this acceleration is achieved without compromising, and even slightly improving, the rate-distortion performance. 
Our work makes GMM-based codecs practically viable, paving the way for their broader adoption in high-performance image compression. 

\bibliography{gmm.bib} 
\bibliographystyle{ieeetr}

\end{document}